# Dual-grating single-shot pump-probe technique

Tianchen Yu [a], Junyi Yang [a,*], Wenfa Zhou [b], Zhongguo Li [c], Xingzhi Wu [d], Yu Fang [d], Yong Yang [e,*], Yinglin Song [a,*]

[a]School of Physical Science and Technology, Soochow University, Suzhou 215006, China

[b]Department of Physics, Harbin Institute of Technology, Harbin, 150001, China

[c]School of Electronic and Information Engineering, Changshu Institute of Technology, Changshu 215500, China

[d]Jiangsu Key Laboratory of Micro and Nano Heat Fluid Flow Technology and Energy Application, School of Physical Science and Technology, Suzhou University of Science and Technology, Suzhou 215009, China

[e]School of Optoelectronic Science and Engineering, Soochow University, Suzhou 215006, China

**\*Corresponding author**

E-mail address: yjy2010@suda.edu.cn (Junyi Yang), yangy@suda.edu.cn (Yong Yang) ylsong@suda.edu.cn (Yinglin Song)

**Abstract**

A simple and effective single-shot pump-probe technique is reported for studying the ultrafast dynamic processes in various materials. Using only two commercial gratings, a large time window of ~ 95.58 ps is spatially encoded in a single probe pulse, and single-shot time-resolved measurements are implemented. This time window exceeds the maximum reported values for single-shot pump-probe techniques using the echelon or angle beam encoding strategy. The phase difference problem in the echelon encoding strategies is also eliminated and a customized echelon is not needed in this technique. The ultrafast dynamic processes of ZnSe and indolium squaraine at a wavelength of 650 nm were investigated using this technique.

## 1. Introduction

It is important to study the ultrafast dynamic processes inside materials for the development of optoelectronic devices, ultrafast optical switches, and solar cells [1-3]. Using the pump-probe technique, many ultrafast phenomena such as multiphoton absorption, charge transfer, and energy transfer, have been studied in depth [4]. However, the traditional pump-probe technique based on the mechanical time delay stage cannot meet the requirements of exploring more irreversible phenomena [5-7]. The single-shot pump-probe technique is a time-resolved measurement tool that can obtain the photoexcitation dynamics of materials with only a single probe pulse. Compared with the traditional pump-probe technique using a mechanical time-delay stage, the single-shot pump-probe technique has the advantages of single-shot diagnosis, in-situ measurement, and a higher acquisition rate [8]. The measurable time window of the single-shot pump-probe technique depends on the time window within which a single probe pulse can be encoded; therefore, achieving both high time resolution and a large time window is a state-of-the-art research area. The spatial encoding strategy is an effective way to realize the single-shot pump-probe technique. In this strategy, the time delay is mapped to different spatial positions of the probe beam, and then the intensity distribution of the probe beam is recorded using a CCD/CMOS camera, thereby realizing single-shot dynamic measurements. At present, the main spatial encoding strategies include the echelon encoding strategy [6, 7, 9-12], angle beam encoding strategy [13-17], and grating encoding strategy [18-21]. As far as we know, the reported maximum time window of the echelon coding strategy is ~ 37.7 ps, and the phase difference of the echelon surface can lead to the image blurring problem, which is the main reason that limits the larger time window for the echelon encoding strategy [11]. The maximum time window reported by the angular beam encoding strategy is ~ 60 ps, and a large sample area (2-3 cm) is required to achieve this time window [17]. The grating encoding strategy uses the angular dispersion of the grating to generate a titled pulse front and therefore realizes the time-delay encoding of a single pulse [20]. Compared to the angular beam encoding strategy, this strategy does not require a large sample area. In our previous work [22], we reported a single-shot pump-probe technique that uses a combination encoding strategy of a commercial grating and a simple echelon to achieve a time window of 109 ps, which effectively eliminates the phase difference problem of the echelon encoding strategy. To the best of our knowledge, this is the first study to achieve a time window of more than 100 ps using a spatial encoding strategy. In addition, we demonstrate that a single commercial grating can encode a time window of ~ 57 ps.

In this work, we propose a single-shot pump-probe technique using a dual-grating encoding strategy. By combining two commercial gratings appropriately, a time window close to 100 ps (~ 95.58 ps) is achieved without an additional customized echelon. In addition, this technique maintains the advantage of eliminating the phase difference problem in echelon encoding strategies. The measurements of the ultrafast dynamic processes of ZnSe and indolium squaraine at a wavelength of 650 nm validated the effectiveness of this technique.

## 2. Experimental setup

The schematic of the experimental layout is shown in Fig. 1. A mode-locked Yb: KGW fiber laser (1030 nm, 190 fs FWHM, 6 kHz) pumped optical parametric amplifier (OPA) was used as the laser source, and the output wavelength was selected to 650 nm in our experiment. The repetition frequency of the laser was adjusted to 20 Hz in the experiment. The output femtosecond (fs) laser pulse was divided into a weaker probe beam and a stronger pump beam by a beam splitter. A laser shutter simultaneously controls the passage of the pump pulses and the trigger of the CCD (not shown in Fig. 1). A half-wave plate (HWP) and polarizers $P_1$ and $P_2$ make the polarizations of the pump and probe beams orthogonal to each other.

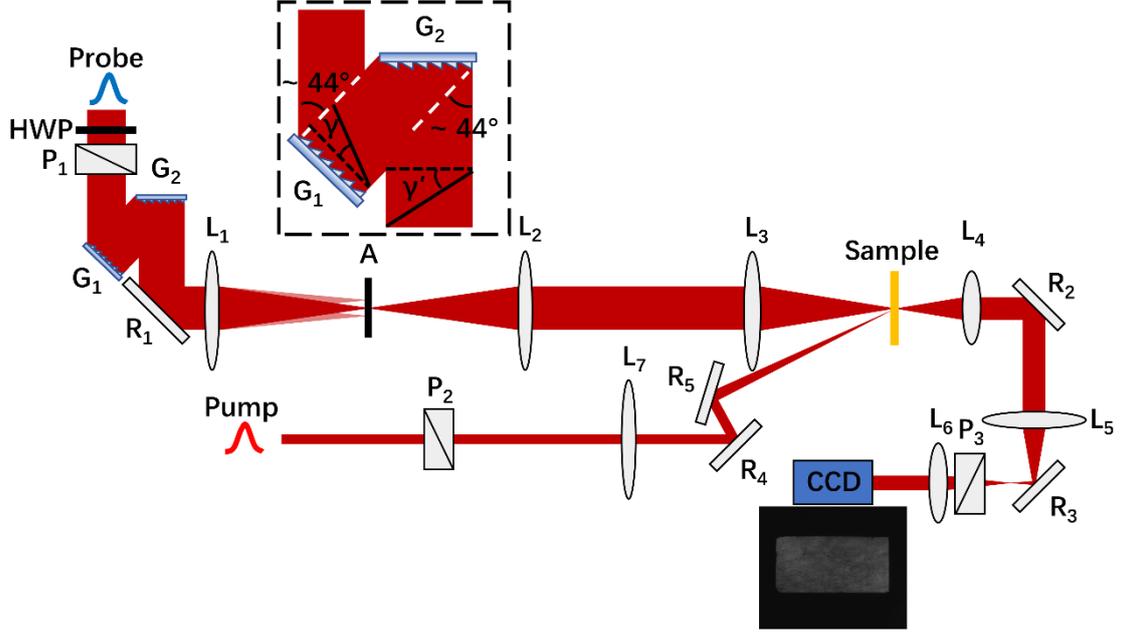

**Fig. 1.** Experimental setup of the single-shot pump-probe technique. HWP: half-wave plate; $G_1$-$G_2$: gratings; $R_1$-$R_5$: mirrors; $L_1$-$L_7$: lenses; $P_1$-$P_3$: polarizers; A: aperture; black dotted box: schematic of dual-grating encoding; CCD: charge-coupled device.

The probe beam is expanded and irradiated on ruled grating $G_1$ at an incident angle of ~ 44°. Since the blazed angle $\theta_b$ of $G_1$ is ~ 22.02°, the geometric reflection beam will be parallel to the grating normal. Due to the angular dispersion of the grating, a tilted pulse front is introduced into the probe pulse, while the phase front is maintained orthogonal to the propagation direction (as shown in the enlarged view inside the dotted box in Fig. 1, the solid line represents the titled pulse front, and the dashed line represents the phase front). The tilt angle $\gamma$ between the pulse front and the phase front can be expressed as [20, 23]:

$$\tan \gamma = \omega \left| \frac{d\beta}{d\omega} \right| = \sin(2\theta_b) \tag{1}$$

where $\omega$ is the center frequency of the laser pulse and $\beta$ is the exit angle of the beam relative to the grating normal. Using Eq. (1), we can determine the tilt angle $\gamma = 34.8°$ in our experiment. This corresponds to an encoded time delay $T_1$ along the grating length, which can be expressed as:

$$T_1 = \frac{\tan \gamma \cdot L_{eff}}{c} = \frac{L_{eff} \cdot \sin(2\theta_b)}{c} \tag{2}$$

where $L_{eff}$ is the effective length of the grating and $c$ is the light speed. The size of the grating used in the experiment was 25×12.5 mm², and the ruled area was 24×11.5 mm² with 75 grooves per mm. Therefore, $L_{eff}$ is 24 mm, and the encoded time delay $T_1$ is 55.6 ps. The encoded time step $\Delta t$ can also be determined by the groove width $d$:

$$\Delta t = \frac{d \cdot \sin(2\theta_b)}{c} \tag{3}$$

Therefore, $\Delta t$ can be calculated as ~ 30.9 fs. The encoded probe pulse is then irradiated on the ruled grating $G_2$ of the same size in the same way (incidence angle ~ 44°). The distance between $G_1$ and $G_2$ is approximately 6 cm. In this way, the probe pulse is further encoded by $G_2$, and the encoded time window is equal to $T_1$. It should be noted that only a part of the probe beam reflected by $G_1$ is projected to $G_2$, so the total encoded time window $T_2$ and the total encoded time step $\Delta t'$ are

$$T_2 = T_1 \cdot \cos(2\theta_b) + T_1 \tag{4}$$

$$\Delta t' = \Delta t \cdot \cos(2\theta_b) + \Delta t \tag{5}$$

Using Eqs. (4) and (5), we can obtain that the total encoding time window $T_2$ and time step $\Delta t'$ are ~ 95.6 ps and 53.1 fs, respectively, in our experiment. Correspondingly, the total titled angle $\gamma'$ can be calculated as:

$$\tan \gamma' = \frac{T_2 \cdot c}{L_{eff}} \tag{6}$$

Thus, the total titled angle $\gamma' = 50.1°$ in our experiment. The probe beam encoded by the dual-grating $G_1$-$G_2$ is spatially filtered by a 4$f$ system $L_1 - L_2$ ($f$ = 200 mm) with an aperture A (diameter ~ 1.5 mm) placed at the focal plane. Then, the probe beam is converged by lens $L_3$ ($f$ = 150 mm) to the sample at the back focal plane and relayed by lenses $L_4$ ($f$ = 50 mm), $L_5$ ($f$ = 100 mm), and $L_6$ ($f$ = 80 mm) to the CCD, which is placed on the image plane of $G_2$. A crossed polarizer $P_3$ is placed in front of the lens $L_6$ to minimize pump light scattering on the image. The pump beam is converged to the probe spot on the sample by lens $L_7$ ($f$ = 500 mm). The optical path of the pump beam is adjusted by a mechanical time-delay stage (not shown in Fig. 1) to overlap the pump and probe pulses in the time window of the grating. Therefore, the dynamic process of the pump-excited sample will be recorded by the encoded probe pulse and mapped to different spatial positions on the CCD panel. In our dual-grating encoding strategy, since the probe beam is normal to the grating surface after each reflection, an accurate image of the grating surface of $G_2$ can be obtained (as shown in Fig. 1) without the phase difference problem in the echelon encoding strategy, which allows it to achieve a large time window.

## 3. Experimental results and discussion
### 3.1. Ultrafast dynamic measurements of ZnSe

Polycrystalline ZnSe with a thickness of 1 mm was investigated using this technique. In this experiment, the pump and probe energies were 15.35 μJ and 2.7 nJ, respectively, and the spot sizes on the sample were ~ 2.3 mm and ~ 0.8 mm, respectively. ZnSe has a bandgap of 2.7 eV; therefore, an ultrafast two-photon absorption (TPA) process should be observed at a pump wavelength of 650 nm (1.91 eV). A "pump on" image acquired by the CCD is shown in Fig. 2(a), which is an accurate surface image of $G_2$. A strip region with reduced gray values appears, as indicated by the red arrow in Fig. 2(a). This is due to the ultrafast TPA process in ZnSe. We can define the corresponding position of the time-delay stage as $\Delta z = 0$ mm and translate it to increase the optical path of the pump beam. Images at $\Delta z = 6$ mm and $\Delta z = 11.5$ mm are shown in Fig. 2(c) and 2(e), respectively. The TPA moves from right to left along the grating length and corresponds to time delay shifts of 40 ps and 76.7 ps, respectively, as indicated by the red arrow. Therefore, the time resolution of each pixel is ~ 51.38 fs/pixel.

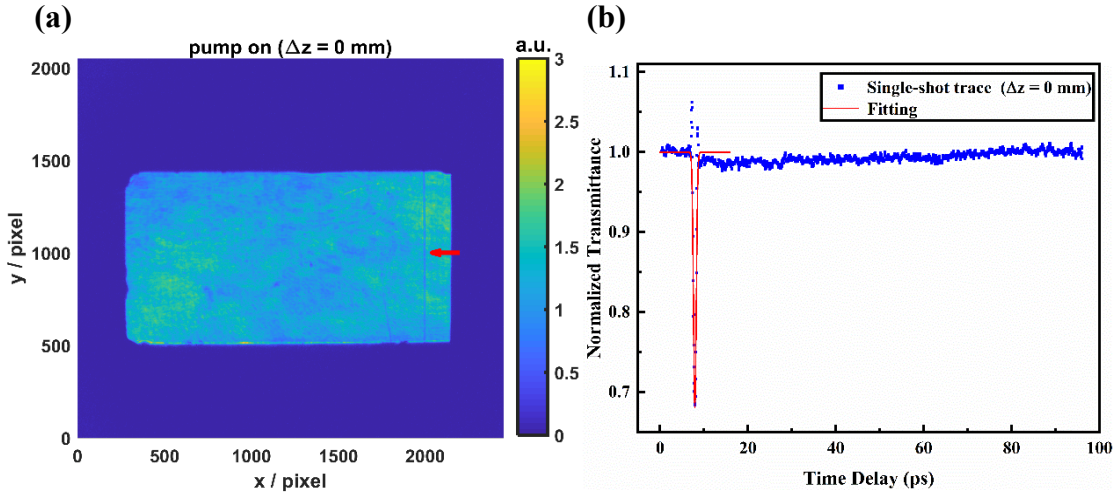

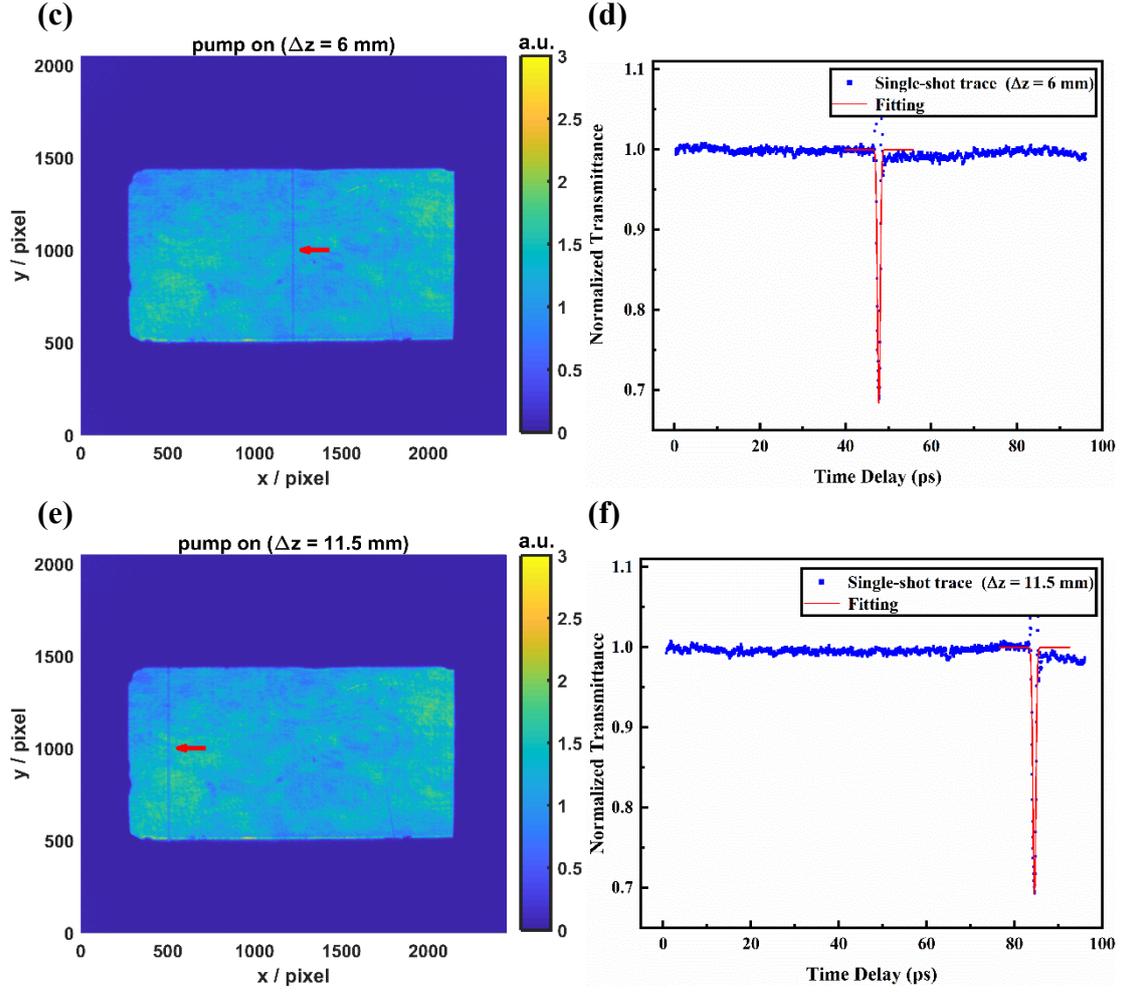

**Fig. 2.** Single-shot measurements of ZnSe. (a) "Pump on" image of $G_2$ at $\Delta z = 0$ mm; (b) extracted single-shot dynamic trace of ZnSe at $\Delta z = 0$ mm; (c) "pump on" image of $G_2$ at $\Delta z = 6$ mm; (d) extracted single-shot dynamic trace of ZnSe at $\Delta z = 6$ mm; (e) "pump on" image of $G_2$ at $\Delta z = 11.5$ mm; (f) extracted single-shot dynamic trace of ZnSe at $\Delta z = 11.5$ mm. Red solid lines: Theoretical fitting.

To extract the dynamic traces of ZnSe, the pixels along the grating width (groove direction) were integrated and then normalized using the following equation:

$$T_{nor} = \frac{I_{11} - I_{10}}{I_{01} - I_{00}} \tag{7}$$

Here, $I_{ij}$ represents the integrated intensity along the groove direction. For the first subscript, "$i = 1$" represents the "pump on" condition, and "$i = 0$" represents the "pump off" condition. Similarly, for the second subscript, "$j = 1$" represents the "probe on" condition, and "$j = 0$" represents the "probe off" condition. The pixels along the grating length were converted into time delays, and the extracted single-shot dynamic traces of ZnSe are shown in Fig. 2(b), 2(d), and 2(f). The total time window is ~ 95.58 ps, which is in good agreement with the value calculated by Eq. (3). Over the entire time window, the most significant feature of the single-shot dynamics of ZnSe is the emergence of a sharp TPA valley. The sharp TPA valley represents the cross-correlation between the pump and probe beams, and it moves along the long-delay direction as the optical path of the pump increases. The TPA valley can be fitted by a Gaussian profile with a full width at half maximum (FWHM) of 776 ± 21 fs, as shown by the red lines in Fig. 2(b), 2(d), and 2(f). This value is greater than the autocorrelation width of ~ 269 fs (FWHM) of the output laser pulse and can be attributed to pulse broadening due to the angular spectral dispersion of the grating, as well as time blurring by the angular crossing of the pump and probe beams [22]. In addition, there are small upward spikes on either side of the TPA valley, which may be related to the two-beam coupling of the chirped pulse due to the nonlinear refractive coefficient of the material [24, 25], which was also observed in previous TPA measurements at a wavelength of 515 nm [22]. After the ultrafast TPA process was completed, the transmittance of ZnSe did not fully recover to the initial value, as can be observed in Fig. 2(b), 2(d), and 2(f). A normalized transmittance decrease of approximately 0.01 was maintained until the end of the entire time window. This is attributed to the TPA-induced carrier absorption process, which is an equivalent fifth-order nonlinear process compared to the TPA process (third-order nonlinear process); therefore, it is generally much weaker than the TPA process. The lifetime of the carriers in ZnSe is usually on the order of the nanosecond (ns) regime [26]; therefore, no obvious carrier recovery characteristics were observed in our time window.

**3.2. Ultrafast dynamic measurements of an indolium squaraine**

An indolium squaraine (ISQ)/DMF solution ($5.05 \times 10^{-5}$ mol/L) was also investigated using this technique. ISQ has a strong linear absorption band in the wavelength range of 600 - 700 nm, which is attributed to the transition from the HOMO to the LUMO [27]. Under the same experimental conditions as ZnSe (except for a 650 nm bandpass filter placed in front of the CCD to filter out possible fluorescence interference), the single-shot dynamic process of ISQ was measured. The acquired grating image is shown in Fig. 3(a). A region of significant light intensity increase appears on the grating image. This abrupt change in light intensity occurs when the pump pulse acts on the sample. Due to the strong linear absorption of ISQ, the particles in the ground state $S_0$ are rapidly depopulated and transition to the excited state $S_n$, resulting in a significant decrease in the absorption of the probe light, i.e., the ground state bleaching (GSB). The corresponding extracted single-shot dynamic trace is shown in Fig. 3(b). An average dynamic trace with 10 laser shots is also given in Fig. 3(b) with a significantly improved signal-to-noise ratio. The improvement in the signal-to-noise ratio is because the average of multiple pulses effectively suppresses the effect of energy fluctuations from pulse to pulse. As can be seen, the transmittance of the probe light increases dramatically over the action time of the pump pulse, and no obvious recovery of the signal occurs within a 100 ps time window. Conversely, a slight and slow enhancement in the transmittance signal emerged as the time delay increased. In addition, an inflection point occurs at approximately 80 ps, after which the signal begins to wane. This trend is consistent with the results reported in ref [27] by a conventional pump-probe technique at a wavelength of 532 nm, which can be attributed to the transition from the $S_n$ state to the $S_1$ state.

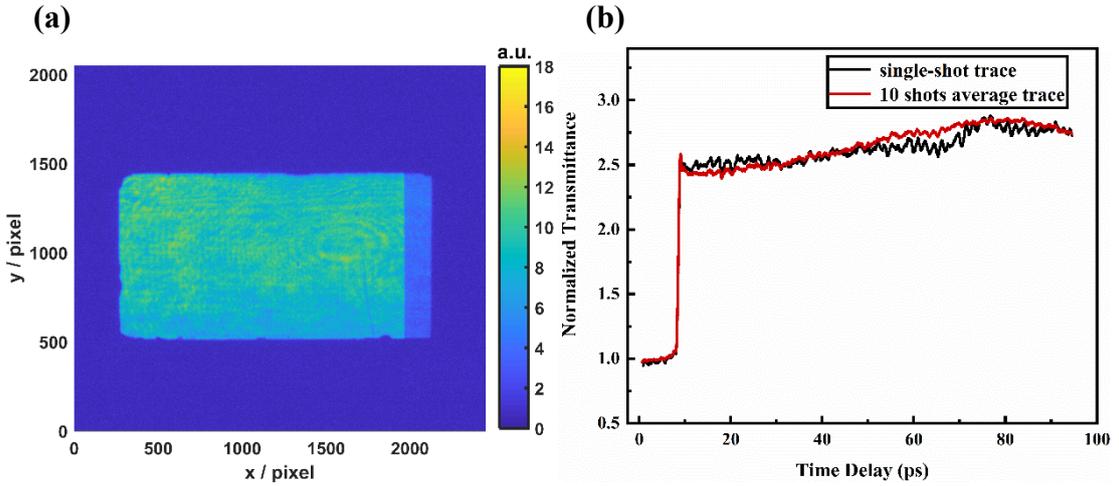

**Fig. 3.** Single-shot measurements of ISQ. (a) "Pump on" image of $G_2$; (b) extracted dynamic traces of ISQ. Solid black line: single-shot dynamic trace; solid red line: average dynamic trace with 10 laser shots.

### 3.3. Comparison of the dual-grating encoding strategy and the echelon and grating combination encoding strategy

Studies of the different dynamic processes of ZnSe and ISQ have successfully verified the effectiveness of our single-shot pump-probe technique. Here, we compare the dual-grating (DG) encoding strategy with the previously reported echelon and grating combination (EG) encoding strategy [22], both of which can achieve a large time window of ~ 100 ps. The principles of the two encoding strategies are shown in Fig. 4.

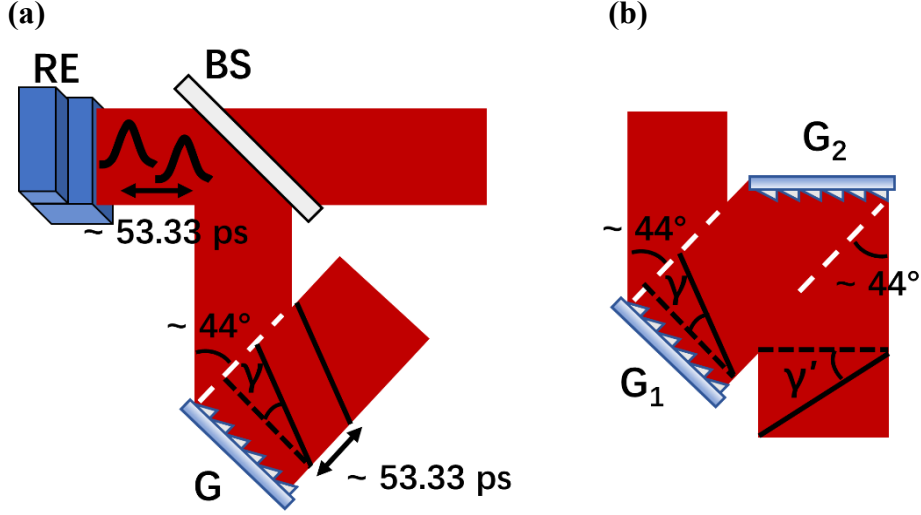

**Fig. 4.** Principles of the EG and DG encoding strategies. (a) EG encoding strategy [22]; (b) DG encoding strategy.

**Table 1**. Comparison of the parameters of the EG and DG encoding strategies.

| Encoding strategy | Time step | Measured time resolution | Encoded time window | Spatial encoding type | Reference |
|---|---|---|---|---|---|
| echelon and grating combination (EG) | 30.9 fs | ~ 855 fs | 109 ps | 2D spatial encoding | [22] |
| Dual-grating (DG) | 53.1 fs | ~ 776 fs | 95.6 ps | 1D spatial encoding | This work |

For the EG encoding strategy in Fig. 4(a), the probe beam is first illuminated on a two-step reflective echelon (RE). Due to the height difference of the steps (8 mm in ref [22]), the reflected probe pulse is split vertically into two subpulses with a time interval of ~ 53.33 ps. The two subpulses are then projected onto the grating surface by a beam splitter BS, and at an incidence angle of ~ 44°, the two subpulses are spatially encoded in the same way as discussed in **Section 2**, along the grating length (horizontal direction). Therefore, two parallel subpulses with the same pulse front tilt $\gamma = 34.8°$ (time window of ~ 55.6 ps) and a time separation of ~ 53.33 ps are obtained, and this is a two-dimensional (2D) spatial encoding. Overall, a total time window of ~ 109 ps can be obtained using the EG encoding strategy. Comparatively, as discussed in **Section 2**, the DG encoding strategy in Fig. 4(b) utilizes the same spatial encoding strategy twice by the gratings and obtains a total time window of ~ 95.58 ps. Although the time window obtained by the DG encoding strategy is slightly smaller than that of the EG encoding strategy, this is a one-dimensional (1D) spatial encoding. A comparison of the parameters of the two encoding strategies is summarized in Table 1. There are several advantages of the DG encoding strategy: i) a customized echelon is not needed; only by combining commercial gratings can a large time window close to 100 ps be obtained. Therefore, the DG encoding strategy is easier to implement than the EG encoding strategy; ii) the DG encoding strategy is a 1D spatial encoding that always encodes the time delay along the grating length. In this way, similar to the echelon [7, 10] and angle beam encoding strategies [13, 14, 17], spectral decomposition can be carried out in the spatial dimension orthogonal to the grating length. In contrast, due to the use of 2D spatial encoding, there is a conflict between the dimensions of spectral decomposition and spatial encoding in the EG encoding strategy, which makes it difficult to adopt a similar spectral decomposition method. Therefore, the DG encoding strategy is more suitable for achieving a time-wavelength resolved single-shot 2D spectrum, i.e., single-shot transient absorption spectroscopy, with a large time window, and this is the direction of our future work. In addition, in this work, although we only demonstrate a single-shot pump-probe technique using dual gratings, a larger time window can be achieved by simply increasing the combination of gratings. Therefore, we believe that the DG encoding strategy is more effective than the EG encoding strategy.

## 4. Conclusion

In summary, we implemented a single-shot pump-probe technique using a dual-grating spatial encoding strategy, and a large time window of ~ 95.58 ps was obtained, which exceeds the maximum time window reported for the echelon or angle beam encoding strategy. Maintaining accurate imaging of the grating surface also eliminates the phase difference problem that exists in the echelon encoding strategy. Using this technique, we studied the single-shot dynamic process of ZnSe, and the results showed that ZnSe has an ultrafast TPA process followed by a weaker carrier absorption process at a wavelength of 650 nm. An intense GSB process in an indolium squaraine (ISQ)/DMF solution was also observed using this single-shot technique under the same experimental conditions. A comparison with the previously reported echelon and grating combination encoding strategy highlights the advantages of this encoding strategy. Therefore, based on the dual-grating (DG) encoding strategy, a simple and effective single-shot pump-probe technique is implemented, and we believe this technique will be a promising tool for the single-shot diagnosis of ultrafast dynamics in various materials.

**Acknowledgments**

We gratefully acknowledge the National Natural Science Foundation of China (11704273, 51607119), National Safety Academic Fund (U1630103), and Science and Technology Innovation Team of Guizhou Education Department (Grant No. [2023]094).

**Conflict of interest**

The authors declare no conflicts of interest.